# Calculating Web Impact Factor for University Websites of Jammu and Kashmir: a Study

Muneer Ahmad[1], Dr. M. Sadik Batcha[2], Wasim Rashid[3], Obaid Hafiz[4]

[1]Ph.D Research Scholar, Department of Library and Information Science,
Annamalai University, Tamil Nadu, (India)

[2]Associate Professor, Department of Library and Information Science,
Annamalai University, Tamil Nadu, (India)

[3]Library Professional, Central University of Kashmir, Nowgam Srinagar, Jammu & Kashmir, India

[4]Library Professional, Central University of Kashmir, Nowgam Srinagar, Jammu & Kashmir, India

**ABSTRACT**

*This paper examines and explores the web impact factor through a webometric study of the present 12 University Websites of Jammu and Kashmir. Identifies the domain systems of the websites; analyzes the number of web pages and link pages, and calculates the External Link WIF or simple web impact factor (WIF) and external web impact factor of all the University websites. Also reflects that some university websites have higher number of web pages, but correspondingly their link pages are very small in number and websites fall behind in their simple and external link web impact factor. It found that the Cluster University of Jammu ranked 1(0.9018)in Internal Link WIF of Websites in Jammu and Kashmir. Shri Mata Vaishno Devi University ranked 1 (0.7249) in External Link Web Impact Factor.*

## I.INTRODUCTION

Webometric is an attractive and one of the most relevant info-metric sub-discipline among traditional bibliometrics approaches. The availability of reliable web data is important to achieve standard results, despite of tool related technical problems and intermediaries requires for web data recovery. The history of library and information science has seen the extensive development of different types of metrics studies. Two major breakthroughs in the history of information were the advent of the printing press in 1450 and the creation of the Internet in 1990, both of which resulted in information explosions. In the digital age, university websites are very important to their stakeholders and there is a need to assess their ranking. Webometric is the study of the quantitative aspects of the construction and use of information resources, structures and technologies on the web, drawing on bibliometrics and informetric approaches. It covers research of all network based communication using quantitative measure. Webometrics, in future, may become one of the most interesting research areas for the vast collection of electronic information available on the publicly indexable web. Paisley rightly identified this area as the future domain of bibliometrics research. The World Wide Web has now become one of the main sources of information on academic and research activities and therefore it is an excellent platform to test new methods of evaluating webometric activities. Webometric studies have focused their analyses mainly in the performance of the academic web domains, because academic institutions like colleges and universities are stable and well-defined institutions on the Web since long time ago. Furthermore, this interest is due to the possibility of building web indicators which explain the academic activity and production (Scharnhorst and Wouters, 2006[1]; Thelwall, 2001, 2002)[2]. This has produced several studies that compare the academic web development in different countries through web indicators (Ingwersen, 1998[3] or through visualization (Heimeriks and Van Den Besselaar, 2006)[4]

**WEB IMPACT FACTOR (WIF)**

*Calculating Revised Web Impact Factor (RWIF)*

The WIF is generally defined as the ratio between the number of links received and the total number of web pages of a particular website (Idrees, 2015)[5]. There are three types of link namely, outlinks or external links which are HTML code on the website which allows site visitors to access other websites, inlinks or backlinks which are hyperlinks on someone else's website that direct visitors to your site, and self-links which are navigational links used in a website to direct users from one page to another page within the site (Noruzi, 2006)[6]. Therefore, according to the Noruzi, there are three types of WIF including Overall WIF, Inlink (Revised) WIF, and Self-link WIF (Noruzi, 2006)[6]. The formula for calculating Overall WIF is:

**Overall Web Impact Factor = X+Y+Z/P**

Where, the variable X represents the number of outlinks, Y represents the total backlink, Z represents the total self-link count and P represents the number of web pages published on the websites which are indexed by a search engine.

However, Noruzi's research shown that self-links for the website under evaluation can provide ambiguous results, as the number of self-links can be manipulated by different means by site owners. For example, in some cases, self-link counts increase because of email addresses associated with websites, which is identified by the search engine as links to that specific domain. On the other hand, self-links are an important percentage of the links that a website receives; therefore, the author expressed that self-links are less meaningful than inlinks because self-links within a website can be created for navigation purposes rather than for endorsing the contents (Noruzi, 2006)[6]. In addition, the importance of inlinks is threefold (Vaughan & Thelwall, 2005)[7] namely;

- more visibility on the web and potentially more traffic to the site
- better coverage by search engines, and
- higher ranking in search results

The RWIF is the result of excluding self-links for a website, thus establishing an analogous impact factor (Idrees, 2015)[5]. So in this study, the authors calculated the RWIF for websites in Jammu and Kashmir Universities as shown below:

**Revised Web Impact Factor = X/Y**

Where, the variable X represents the inlinks (external backlinks) to the website and Y represents the number of web pages published on the websites which are indexed by a search engine.

**The details of all the universities are listed in Table 1, which gives a brief sketch of all 12 universities in Jammu and Kashmir with their year of establishment, located city/ town with URL.**

| S.No. | Name of the University | Year of Establishment | Short Name | City/Town | URL |
|---|---|---|---|---|---|
| | **Table 1 – List of Universities in Jammu and Kashmir** | | | | |
| 1 | **Baba Ghulam Shah Badshah University** | 2002 | BGSBU | Rajouri | http://www.bgsbuniversity.org |
| 2 | **Central University of Jammu** | 2009 | CUJ | Jammu | http://www.cujammu.ac.in |
| 3 | **Central University of Kashmir** | 2009 | CUK | Ganderbal | http://www.cukashmir.ac.in |
| 4 | **Cluster University of Jammu** | 2016 | CUJ | Jammu | http://www.clujammu.in |
| 5 | **Cluster University of Srinagar** | 2016 | CUS | Srinagar | http://www.cusrinagar.edu.in |
| 6 | **Islamic University of Science and Technology** | 2005 | IUST | Awantipora | http://www.islamicuniversity.edu.in |
| 7 | **University of Jammu** | 1969 | UOJ | Jammu | http://www.jammuuniversity.in |
| 8 | **University** | 1956 | UOK | Srinagar | http://www.kashmiruniversity.net |
|    | of Kashmir                                                   |      |        |           |                              |
| 9  | **Sher-e-Kashmir University of Agricultural Science & Technology** | 1982 | SKAUST | Srinagar  | http://www.skuastkashmir.ac.in |
| 10 | **Shri Mata Vaishno Devi University**                        | 1999 | SMVD   | Katra     | http://www.smvdu.net.in      |
| 11 | **Sher-e-Kashmir Institute of Medical Science**              | 1977 | SKIMS  | Soura     | http://www.skims.ac.in       |
| 12 | **National Institute of Technology**                         | 1960 | NIT    | Hazratbal | http://www.nitsri.ac.in      |

## II.LITERATURE REVIEW

In 1955, Eugene Garfield introduced the term *impact factor* to measure the overall influence of a journal's articles on later literature, subsequently calling it *journal impact factor* (JIF). JIF is a measure of frequency that reflects the average number of citations received by a journal article within a specified period, normally one year. The JIF serves as a measure of the importance of a journal within its field, as journals with higher JIFs are considered to be more important than those with lower ones. For websites, the impact factor is calculated by the number of hyperlinks divided by the number of web pages for a particular website and is called the *web impact factor* (WIF). For WIF, citation data are substituted by *inlinks* (hyperlinks referring to the website or web page) and the idea of self-citation is replaced by *self-links* (navigational links within a website). Webometric approaches require four types of analysis of a website: content analysis (web page), structure analysis (web links), usage analysis (exploiting log files for users' searching behaviour) and technology analysis (search engine performance and optimization) that covers both the usage and structural aspects of the Web (Montelongo, 2013)[8]. The term WIF was introduced by Ingwersen in 1998 and is based on the same pattern as JIF to measure the impact of websites in a particular field of study (Noruzi, 2006)[6]. WIF is considered to provide a website's relative importance and competitive relationship to other websites in the same field or domain. The WIF is generally defined as the ratio between the number of links received and the total number of web pages of a particular website.

**Objectives of the Study**

The primary objective of this study is to examine the WIF for websites of Jammu and Kashmir. Other specific objectives include:

1. To study about the different types of domain of universities in Jammu and Kashmir
2. To study the growth of Universities websites in Jammu and Kashmir
3. To calculate the Revise Web impact factor (RWIF) and Internal Web Impact Factor for the Jammu and Kashmir university websites
4. To calculate the maximum number of inlinks (external links) that refer to the Jammu and Kashmir Universities' websites under study

This study also undertakes link analysis for the inlinks to websites of the selected Jammu and Kashmir universities. In doing so, the total number of inlinks is calculated, which refers to a particular website anywhere in its text body.

**Scope**

The present study makes a webometric analysis of University websites in Jammu and Kashmir. The study examined the websites of 12 universities in the state and aimed at to establish a kind of academic ranking of these websites by measuring their web impact factor. The ranking of websites will help the reader to compare and identify university websites in Jammu and Kashmir according to their WIF.

**Methodology**

When undertaking a WIF study, it is necessary to select a suitable search engine that will count the number of pages in the web site studied, and the number of pages linking to the web site. It should have a large database, covering as much of the Web as possible**.** Currently, Google satisfies these requirements most fully, with one of the largest databases and search commands both for links and for number of pages at a web site.

The research followed the descriptive approach. The research method used in this study is survey method. In order to collect data, we have used the list of universities in Jammu and Kashmir provided by the University Grant Commission of India (www.ugc.ac.in). Before using the list, we have checked the access of each university web sites. The study included 12 universities in Jammu and Kashmir.

In present study, therefore, the authors decided to choose Google and Yahoo search engines and Small Seo Tool and Seo Chat application to collect data, i.e. number of web pages in a university website, number of external backlinks and outlinks or external links.

**Data Collection Strategy**

The required data were collected in May 2018 using search engine namely Google for retrieving number of web pages and Inlinks or backlinks to the website, Yahoo for retrieving number of web pages and Small Seo Tool and Seo Chat service tools for getting Outlinks or External links in order to limit errors associated with frequent website updates. In each search engine, there were some specific search keywords assigned by the search engines to retrieve the required information from the Web. These specific search keywords along with search syntax have been presented in the following table (Table 2a and Table 2b).

| Table 2a: Search Syntax for retrieving no. of web pages ||
|---|---|
| **Search Engine** | **No. of Web Pages** |
| Google | site: www.cujammu.ac.in |
| Yahoo | site:cujammu.ac.in |

| Table 2b: Search Syntax for retrieving no. of Inlinks or backlinks to the website ||
|---|---|
| Inlinks or backlinks to the website | **Google** |
|  | link: www.cujammu.ac.in-site:cujammu.ac.in |

### III.RESULTS AND DISCUSSION

Table 3 shows the domain-wise distribution of the websites of Jammu and Kashmir Universities. Out of 12 Universities, 5 (41.67%) are having ".ac.in" domain, 2 (16.67%) Universities has ".edu.in", 2(16.67%) has ".in" domain and remaining universities have used different domains, one university used ".org", one university used ".net" and one has ".net. in". It is revealed from the table, that "ac.in" domain has been widely used in the websites of Universities in Jammu and Kashmir.

| Table 3: Domain Wise Distribution of Universities in Jammu and Kashmir ||||
|---|---|---|---|
| S.No. | **Domain Name** | **No. of Universities** | **Percentage** |
| 1 | **.ac.in** | 5 | 41.67 |
| 2 | **.edu.in** | 2 | 16.67 |
| 3 | **.in** | 2 | 16.67 |
| 4 | **.net** | 1 | 8.33 |
| 5 | **.org** | 1 | 8.33 |
| 6 | **.net.in** | 1 | 8.33 |
|  | **Total** | 12 | 100 |

Table 4 shows the distribution of Internal-link Web Impact Factor of websites of universities in Jammu and Kashmir. From the table, it is obvious that Cluster University of Jammu is in the first position with 0.9018 of internal links Web Impact Factor. Shri Mata Vaishno Devi University has the second place with 0.6204, Central University of Jammu is in the third place with 0.5891, Cluster University of Srinagar has the fourth place with 0.2123, and Jammu University occupies the fifth place with 0.0874. Central University of Kashmir, Sher-e-Kashmir Institute of Medical Science, Sher-e-Kashmir University of Agricultural Science & Technology and Kashmir University are having sixth, seventh, eighth and ninth positions respectively. No Data available for the Baba Ghulam Shah Badshah University, National Institute of Technology & Islamic University of Science and Technology in the respective sources.

| Table 4: Distribution of Internal Link WIF of Websites of State Universities in Kashmir | | | | | | | | |
|---|---|---|---|---|---|---|---|---|
| S. No. | Name of University | Outlinks or External Links or Outbound Links Count | | | Page Count | | | Revised Web Impact Factor X/Y | Rank |
| | | Small Seo tool | Seo Chat | Average X | Google | Yahoo | Average Y | | |
| 1 | Cluster University of Jammu | 65 | 82 | 73.5 | 114 | 49 | 81.5 | 0.9018 | 1 |
| 2 | Shri Mata Vaishno Devi University | 235 | 255 | 245 | 282 | 22 | 152 | 0.6204 | 2 |
| 3 | Central University of Jammu | 260 | 259 | 259.5 | 636 | 245 | 440.5 | 0.5891 | 3 |
| 4 | Cluster University of Srinagar | 62 | 24 | 43 | 295 | 110 | 202.5 | 0.2123 | 4 |
| 5 | Jammu University | 397 | 396 | 396.5 | 7180 | 1890 | 4535 | 0.0874 | 5 |
| 6 | Central University of Kashmir | 140 | 138 | 139 | 4200 | 1030 | 2615 | 0.0532 | 6 |
| 7 | Sher-e-Kashmir Institute of Medical Science | 212 | 202 | 207 | 7410 | 405 | 3907.5 | 0.0529 | 7 |
| 8 | Sher-e-Kashmir | 24 | 140 | 82 | 752 | 925 | 1677 | 0.0489 | 8 |

| | University of Agricultural Science & Technology | | | | | | | | |
|---|---|---|---|---|---|---|---|---|---|
| 9 | **Kashmir University** | 193 | 122 | 157.5 | 6380 | 1820 | 4100 | 0.0384 | 9 |
| 10 | **Baba Ghulam Shah Badshah University** | 0 | 0 | 0 | 136 | 20 | 78 | | |
| 11 | **National Institute of Technology** | 0 | 1 | | 1 | 3 | 2 | - | |
| 12 | **Islamic University of Science and Technology** | 0 | 0 | 0 | 1 | 1 | 1 | | |

Table 5 illustrates the distribution of External Link Web Impact Factor of university websites of Jammu and Kashmir. From the table it is revealed that Shri Mata Vaishno Devi University stands first with 0.7249 of External Web Impact Factor, the second position is secured by Central University of Jammu and its External Web Impact Factor is 0.5346 followed by Sher-e-Kashmir Institute of Medical Science with 0.1146. Central University of Kashmir has the fourth place with 0.0697. Kashmir University, Cluster University of Jammu, Cluster University of Srinagar, Jammu University, Baba Ghulam Shah Badshah University, and Sher-e-Kashmir University of Agricultural Science & Technology occupy rest of the ranks respectively. No Data available for the Islamic University of Science and Technology and National Institute of Technology in the respective sources.

| Table 5: Distribution of External Link WIF of Websites of State Universities of Jammu and Kashmir | | | | | |
|---|---|---|---|---|---|
| **S.No.** | **Name of University** | **Google Inlinks or Backlinks or Inbound Links Count** | **Google Page Count** | **Web Impact Factor** | **Rank** |
| 1 | **Shri Mata** | 389 | 282 | 0.7249 | 1 |

|  | Vaishno Devi University |  |  |  |  |
|---|---|---|---|---|---|
| 2 | **Central University of Jammu** | 34 | 636 | 0.5346 | 2 |
| 3 | **Sher-e-Kashmir Institute of Medical Science** | 849 | 7410 | 0.1146 | 3 |
| 4 | **Central University of Kashmir** | 293 | 4200 | 0.0697 | 4 |
| 5 | **Kashmir University** | 405 | 6380 | 0.0635 | 5 |
| 6 | **Cluster University of Jammu** | 7 | 114 | 0.0614 | 6 |
| 7 | **Cluster University of Srinagar** | 8 | 295 | 0.0271 | 7 |
| 8 | **Jammu University** | 155 | 7180 | 0.0216 | 8 |
| 9 | **Baba Ghulam Shah Badshah University** | 1 | 136 | 0.0073 | 9 |
| 10 | **Sher-e-Kashmir University of Agricultural Science & Technology** | 3 | 752 | 0.0039 | 10 |
| 11 | **Islamic University of Science and Technology** | 1 | 1 | - | - |

| 12 | National Institute of Technology | 1 | 1 | - | - |

**Findings**

Web Impact Factor and link analysis of Universities websites of Jammu and Kashmir is an unexplored area of webometric research. The present study, hopefully, provides a fair idea and information about the website of all the 12 university websites of J & K. There is a scope for further webometric research in this area. The followings are the major findings of the present study:

- Majority of the Universities 5 (41.67) are having .ac.in domain name for their websites. .ac.in is specially using for academic purpose in India.
- The 4 universities each 2 (16.67) is having .edu.in and .in domain for their university websites.
- Cluster University of Jammu ranked 1 (0.9018) in Internal Link web impact factor of university websites in Jammu and Kashmir.
- Shri Mata Vaishno Devi University ranked first (0.7249) in External Link Web Impact Factor.

**IV. CONCLUSION**

WIF is becoming a more reliable indicator worldwide to measure the scientific utility of websites. Frequent studies have been conducted to calculate different types of WIFs for websites, especially academic websites. This study concludes that WIF, being a quality indicator, helps in measuring the utility of a website, rather than measuring its overall impact on the Web. However, website evaluation methods, using Webometrics tools, require a multidimensional approach and should never be confined to link analysis only. Bibliometric methods on the Web are highly affected by the distributed and diverse nature of the Web (Noruzi, 2006)[6]. Traditional methods of publishing scientific and research material, language barriers, no links to institutional research material on its website, unawareness of Web ranking by webmasters and university Web developers, structural problems in Web designing, limitation of access to material, absence and non-linkage of institutional repositories on the Web are some commonly identified factors that contribute to increase or decrease the impact of an academic website globally.

**REFERENCES**

[1]     A. Scharnhorst and P. Wouters, "Web indicators – a new generation of S & T indicators ?," *Cybermetrics*, vol. 10, no. January, 2006.

[2]     M. Thelwall, "A comparison of sources of links for academic Web impact factor calculations," *J. Doc.*, vol. 58, no. 1, 2002, pp. 66–78.

[3]     L. Björneborn and P. Ingwersen, "Toward a Basic Framework for Webometrics," *J. Am. Soc. Inf. Sci. Technol.*, vol. 55, no. 14, 2004, pp. 1216–1227.

[4]     G. Heimeriks and P. Van Den Besselaar, "Analyzing hyperlinks networks: The meaning of hyperlink

based indicators of knowledge production," *Cybermetrics*, vol. 10, no. 1, 2006.

[5] A. K. Idrees, "Calculating Web impact factor for university websites of Pakistan," *Electron. Libr.*, vol. 33, no. 5, 2015, pp. 883–895.

[6] A. Noruzi, "The Web Impact Factor : A Critical Review *," *Electron. Libr.*, vol. 24. , 2006, pp. 1–10.

[7] L. Vaughan and K. Hysen, "Relationship between links to journal Web sites and impact factors," vol. 54, no. 6, 2002, pp. 356–361.

[8] J. A. Montelongo and A. C. Hernández, "The teachers' choices cognate database for K-3 teachers of Latino English learners," *Read. Teach.*, vol. 67, no. 3, 2013, pp. 187–192.